\documentclass[smallextended]{svjour3}       
\smartqed  
\usepackage{graphicx}
\usepackage{color}
\begin{document}

\title{Observation of pentaquark states and perspectives of further studies}


\author{Cheng-Ping Shen         \and
        Chang-Zheng Yuan 
}


\institute{Cheng-Ping Shen \at
           School of Physics and Nuclear Energy Engineering,\\
           Beihang University, Beijing, China \\
           \email{shencp@buaa.edu.cn}           
           \and
           Chang-Zheng Yuan \at
           Institute of High Energy Physics, \\
           Chinese Academy of Sciences, Beijing, China
}

\date{Received: date / Accepted: date}

\maketitle


The quark model based on quark-antiquark mesons and three-quark
baryons has been very successful in classifying the hadrons.
However, quantum chromodynamics (QCD) does not forbid the
existence of exotic hadronic states with other quark-gluon configurations,
such as glueballs (with no quark), hybrids (with quarks and
excited gluon), multi-quark states (with more than three quarks),
and hadron molecules (bound state of two or more hadrons). It is a
long history of searching for all these kinds of states, however,
no solid conclusion was reached until the recent discovery of
tetraquark and pentaquark states.

The first strong experimental evidence for a pentaquark state,
referred to as the $\mathrm{\Theta}(1540)^+$, was reported in the
reaction $\gamma n \to n K^+ K^-$ in the LEPS
experiment~\cite{leps}. It was a candidate pentaquark state of
$uudd\bar{s}$. However, it disappeared in a larger statistics data
sample in the same experiment and was most probably not a genuine
state~\cite{mao}. The overwhelming studies in theory and in
experiments during that period of time was well accounted in the
statement that ``The story of pentaquark shows how poorly we
understand QCD''~\cite{wilczek_2005}.

A charged charmoniumlike state $Z_c(3900)$ was observed by
BESIII~\cite{zc3900} and Belle~\cite{belley_new} experiments in
2013 in $J/\psi \pi^{\pm}$ system of $e^+e^-\to \pi^+ \pi^- J/\psi$ at
center-of-mass energies around 4.26~GeV. Besides the $Z_c(3900)$,
BESIII and Belle also observed a series of charged $Z_c$ states
including $Z_c(4020)$, $Z_c(4200)$, and $Z_c(4430)$. These states
seem to indicate that a new class of hadrons has been
observed~\cite{liu}. As there are at least four quarks within
these $Z_c$ states, they have been interpreted either as
tetraquark states with a pair of charm-anticharm quarks and a pair
of light quarks, molecular states of two charmed mesons,
or other configurations.

\begin{figure}
\begin{center}
\includegraphics[width=0.5\textwidth]{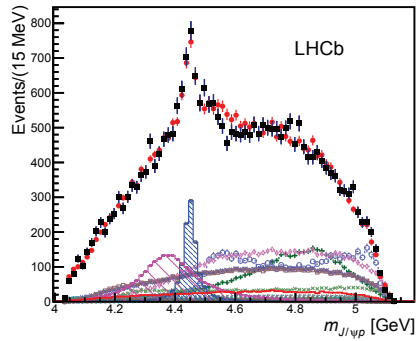}
\end{center}
\caption{Distribution of the $J/\psi p$ invariant mass and the fit
with two $P_c^+$ states~\cite{lhcb}. The purple
hatched histogram is the $P_c(4380)^+$, and the blue hatched
histogram is the $P_c(4450)^+$ signal.}
\label{fig1}       
\end{figure}

Very recently, LHCb experiment reported the observation of two
exotic structures, denoted as $P_c(4380)^+$ and $P_c(4450)^+$, in
the $J/\psi p$ system in $\Lambda_b^0 \to J/\psi K^-
p$~\cite{lhcb}. The $P_c(4380)^+$ has a mass of $(4380\pm 8\pm
29)$~MeV/$c^2$ and a width of $(205\pm 18\pm 86)$~MeV, while the
$P_c(4450)^+$ is much narrower with a mass of $(4449.8\pm 1.7\pm
2.5)$~MeV/$c^2$ and a width of $(39\pm 5\pm 19)$~MeV.
Figure~\ref{fig1} shows the fit to the $J/\psi p$
invariant mass distribution and the $P_c(4380)^+$ (purple hatched histogram)
and $P_c(4450)^+$ (the blue hatched histogram) contributions. Since the valence
structure of $J/\psi p$ is $c\bar{c}uud$, the newly discovered
particles consist of at least five quarks.

Several theoretical interpretations of these states have been
developed, including a pentaquark doublet, hadronic molecules
composed of an anticharm meson and a charm baryon, a $\chi_{c1} p$
resonance for the $P_c(4450)^+$, and so on.

To confirm these pentaquark states, further experimental research
should be pursued with the current available and the forthcoming
experimental data. There have been many suggestions on the
discovery channels for these and other exotic pentaqaurks,
such as (1) in $B$ decays: $B^0 \to p \bar{p} K^0$, $\bar{B}^0\to
D^0 p \bar{p}$; (2) in baryon decays: $\Lambda_c^+ \to p K^0
\bar{K}^0$, $\Lambda_b^0 \to K^- \chi_{c1} p$; (3) in quarkonium
decays: $\Upsilon(nS)\to J/\psi p+X$, $\chi_{cJ} p+X$, and $D^{(*)-} p
+ X$ ($n=1$, 2, 3); (4) in $e^+e^-$ continuum
process: $e^+e^- \to J/\psi p + X$,  $\chi_{cJ} p+X$,  $D^{(*)-} p + X$ and $D^{(*)} \Lambda + X$. It is
clear that a systematic search for baryon-meson resonances should
be pursed in various processes, where the baryon could be $p$,
$\Lambda$, $\Sigma$, $\Xi$, $\Omega$, $\Sigma_c$, ..., and the
meson be $\pi$, $\eta$, $\omega$, $\phi$, $K$, $D$, $J/\psi$,
$\chi_{cJ}$, and so on~\cite{ref1}~\cite{ref2}.

The Belle-II experiment is going to take data in 2018, the maximum
energy that the Super-KEKB can reach is 11.24~GeV, which is just
above the $\Lambda_b$-pair mass threshold. The $P_c(4380)^+$ and
$P_c(4450)^+$ can be cross checked in the same $\Lambda_b$ decays
if the production cross section of $\Lambda_b$ is large enough.

It is worth pointing out that the tetraquark and pentaquark
candidates mentioned above have a pair of charm-anticharm quarks
which may annihilate. Observations of states like $T_{cc}^+$ ($cc\overline{ud}$)
or $\mathrm{\Theta}_c^0$ ($uudd\bar{c}$) or $P_{cc}^0$
($ccdd\bar{u}$) or similar serve as better evidences for multiquark
states.

In summary, with the observation of the candidate tetraquark and
pentaquark states, a new hadron spectroscopy is being revealed.
Further studies along this line may strengthen our understanding
of how strong interaction works at low energy and thus a better
understanding of the matters around us.


\end{document}